\documentclass[conference]{IEEEtran}
\IEEEoverridecommandlockouts
\usepackage{cite}
\usepackage{amsmath,amssymb,amsfonts}
\usepackage{algorithmic}
\usepackage{graphicx}
\usepackage{textcomp}
\usepackage{enumitem}
\usepackage{mdframed}
\usepackage{xcolor}
\usepackage{url}
\def\BibTeX{{\rm B\kern-.05em{\sc i\kern-.025em b}\kern-.08em
    T\kern-.1667em\lower.7ex\hbox{E}\kern-.125emX}}
\begin{document}

\title{Bridging eResearch Infrastructure and Experimental Materials Science Process in the Quantum Data Hub\\
\thanks{Partially funded by NSF grants 1909875, 1906325 and	1906383.}
}

\author{\IEEEauthorblockN{Amarnath Gupta}
\IEEEauthorblockA{\textit{San Diego Supercomputer Center} \\
\textit{University of California San Diego}\\
La Jolla, CA, USA \\
a1gupta@ucsd.edu}
\and
\IEEEauthorblockN{Shweta Purawat}
\IEEEauthorblockA{\textit{San Diego Supercomputer Center} \\
\textit{University of California San Diego}\\
La Jolla, CA, USA \\
shpurawat@ucsd.edu \\}
\and
\IEEEauthorblockN{Subhasis Dasgupta}
\IEEEauthorblockA{\textit{San Diego Supercomputer Center} \\
\textit{University of California San Diego}\\
La Jolla, CA, USA \\
sudasgupta@ucsd.edu }
\and
\IEEEauthorblockN{Pratyush Karmakar}
\IEEEauthorblockA{\textit{San Diego Supercomputer Center} \\
\textit{University of California San Diego}\\
La Jolla, CA, USA \\
pkarmakar@ucsd.edu}
\and
\IEEEauthorblockN{Elaine Chi}
\IEEEauthorblockA{\textit{San Diego Supercomputer Center} \\
\textit{University of California San Diego}\\
La Jolla, CA, USA \\
ychi@ucsd.edu}
\and
\IEEEauthorblockN{Ilkay Altintas}
\IEEEauthorblockA{\textit{San Diego Supercomputer Center} \\
\textit{University of California San Diego}\\
La Jolla, CA, USA \\
ialtintas@ucsd.edu}
}

\maketitle

\begin{abstract}
Experimental materials science is experiencing significant growth due to automated experimentation and AI techniques. Integrated autonomous platforms are emerging, combining generative models, robotics, simulations, and automated systems for material synthesis. However, two major challenges remain: democratizing access to these technologies and creating accessible infrastructure for under-resourced scientists. This paper introduces the Quantum Data Hub (QDH), a community-accessible research infrastructure aimed at researchers working with quantum materials. QDH integrates with the National Data Platform, adhering to FAIR principles while proposing additional UNIT principles for usability, navigability, interpretability, and timeliness. The QDH facilitates collaboration and extensibility, allowing seamless integration of new researchers, instruments, and data into the system. 
\end{abstract}


\section{Introduction}
\label{sec:intro}
Experimental materials science is witnessing an unprecedented growth spurt in the use of automated experimentation and application of Artificial Intelligence (AI) techniques. The idea of self-driving labs \cite{abolhasani2023rise} builds on the convergence of machine learning, lab automation (synthesis, separation, purification and characterization) and robotics (reagent preparation and sample transfer). Automated materials discovery techniques \cite{suh2020evolving} exploit machine learning to search over published and experimental data for materials that are predicted to satisfy user-specific properties. \cite{lu2023artificial} demonstrates an integrated autonomous platform that combines generative models, robotics, multiphysics simulations, and automated synthesis and characterization systems for 2D material synthesis. As Maqsood et al \cite{maqsood2024future} state, today's infrastructural demands for materials science ``ranges from basic data-fitting techniques to $\ldots$ semi-autonomous experimentation, experimental
design, knowledge generation, hypothesis formulation and the orchestration of specialized AI modules''. 

Despite this advancement and all the excitement around it, there are two fundamental issues that the materials eScience community needs to consider: 1) How can this cutting edge research methodologies be ``democratized'' \cite{abramson2022democratising}?; and  2) How can one design a generic and community-accessible infrastructure that enables under-resourced materials scientists make use of (and contribute to) these advancements? Here, ``democratization'' refers to making resources and infrastructure available to the science researchers who may not be able to own such resources at their own labs or institutions. We use the term ``infrastructure'' to cover every aspect of a modern AI-enabled experimental materials science research process including materials, experimental and characterization procedures, instruments, computational platforms and procedures, and human knowledge in the form of expertise.

This paper presents the Quantum Data Hub\footnote{\url{https://quantumdatahub.sdsc.edu/}} (QDH) as a community-accessible research infrastructure currently targeted for researchers who design, synthesize and characterize quantum materials (e.g., thin-film topological insulators) via instrument-based observation and computation-based methods. This infrastructure, interoperates  with the National Data Platform (\url{nationaldataplatform.org}) designed to provide data and data-related services at the national scale, constituting a suitable e-Science framework for our target science audience. An important aspect of our infrastructure design process rests on the FAIR guiding principle(s) \cite{wilkinson2016fair}. However, we share the viewpoint by Abramson et al \cite{abramson2022democratising} that while FAIR, as guiding principles, encourages robust repeatable science, it does not capture the complexity of the science process, the complex relationships across science data content and the information and resource access needs of scientists. We therefore posit a new additional set of design principles called UNIT that supplements the FAIR principles, and show how these principles are ingrained in the design of the QDH system.

\noindent \textbf{Contributions.} Based on the above context, this paper makes the following contributions:
\begin{enumerate}
 
\item We present the design of the QDH as a Knowledge-Network system that connects people, data, instruments, labs, experimental protocols and AI-based methods. 

\item We introduce UNIT principles for usability, navigability, interpretability, and timeliness to augment FAIR principles when applied within the eResearch process.

\item In alignment with the objectives of QDH to facilitate collaborations among different groups and labs through sharing of data and the material synthesis process, we demonstrate how the extensibility framework of the QDH can be effectively used to include new people, research, instruments and data into the system without any major disruption of the framework, even when a newly onboarded extension is not included in the current information framework.

\end{enumerate}

\section{Related Work}
Material science has made significant advances in the past ten years from theoretical, experimental, and computational viewpoints. The field of computational materials science, which merges computation and material science, has developed significantly due to the major advancements in computational capabilities over time\cite{louie2021discovering}. This relatively new computational paradigm has shown the need for robust, heterogeneous data processing systems that utilize diverse computational units (CPU, GPU, or TPU) and modernized data workflow platforms such as JupyterHub or MLflow \cite{zhu2022materials, butler2018machine, chen2020machine}. Numerous researchers have noted that ``data-driven material discovery" represents the ``fourth industrial revolution," facilitated by the "fourth paradigm of science" \cite{agrawal2016perspective}. In this context, we examined several significant systems and observed that in the recent years, several pioneering efforts have been made to achieve different goals. 

\subsection{Related Platforms that Enable Material Science}
\label{sec:platforms}
A significant class of related platforms can be categorized either as core material science simulations extended with specialized AI/ML techniques tailored to a specific group or as more data-/computation-focused platforms. In this domain, Wang et al. introduced ALKEMIME \cite{wang2021alkemie}, a straightforward server-based functionality that facilitates data integration from various storage systems such as PBS, SLURM, and LTRM, and also offer sophisticated APIs to tailor additional management systems. However, its contributions to the standardization in material science are either very limited or nonexistent, leading to a workflow that is heavily dependent on the ALKEMIME API. This toolkit plays a crucial role in integrating AI with material data. The Materials Simulation Toolkit for Machine Learning (MAST-ML)\cite{jacobs2020materials} aims to reduce human intervention in supervised learning, significantly accelerating the machine learning workflow, although it is not primarily designed for managing data. Since MAST-ML is Open Source and easy to integrate, a significant number of groups have used it, for instance, Lolo \cite{o2016materials}, AFLOW-ML \cite{gossett2018aflow}, matminer \cite{ward2018matminer}, Materials Knowledge Systems in Python project (pyMKS) \cite{brough2017materials}, or MAML\cite{ward2018matminer}. Our evaluation concluded that the variety of tools could be expanded and the capabilities of material informatics tools could be enhanced through the integration of more efficient algorithms. In addition, it is essential to develop a more generalized platform for material science. The AlphaMat effort by Wang et al.\cite{wang2023alphamat} is one of the pioneers in developing such a system. However, none of these systems offer a hosted environment for data science practitioners, nor do they equal the data-centric features of the QDH system, particularly in terms of search capabilities, knowledge graph connectivity, access management, and security protocols.
\subsection{Related Data Systems that Facilitate Collaboration in Material Science}
Several research groups have recognized that material synthesis and other experimental processes can be significantly improved by AI/ML. However, like other sciences, AI/ML is heavily dependent on vast amounts of data. Over the past decade, numerous platforms have emerged, evolving from simple data storage facilities to comprehensive data centers. The Novel Materials Discovery (NOMAD) group\cite{sbailo2022nomad} serves as a prime example of such facilities. It was established in the early 2010s, largely motivated by the US Materials Genome Initiative \cite{mginitiative}. This initiative was launched at a time when several similar projects were already underway, such as AFLOW\cite{curtarolo2012aflowlib}, OQMD\cite{saal2013materials}, and Materials Project\cite{jain2013commentary} each playing a critical role in the storage and management of materials science data.
\newline
The development of NOMAD closely mirrors the functionalities offered by QDH, including the provision of a Jupyter Notebook and a query interface for database interaction. However, QDH stands apart with a few additional features such as a procedure editor, GEMD \cite{GEMD} models and fine-grained access control, which provide users with enhanced control over data and greater flexibility in expressing the material synthesis process using a drag and drop facilities. This paper will provide a detailed discussion of these functionalities. Both platforms are committed to FAIR principles and have gained popularity within the community.
\newline
Additionally, there are several initiatives, such as the Materials Data Facility \cite{blaiszik2019data} and Materials Cloud \cite{talirz2020materials}, dedicated to the open-access storage for different class of materials science data. Moreover, platforms like Zenodo \cite{zenodo2013} generate Digital Object Identifiers (DOIs), assisting researchers in citing material investigations effectively.

\section{Design of the Quantum Data Hub}
\label{sec:design}
The primary vision of the Quantum Data Hub is to capture, manage and support the socio-scientific process of materials synthesis and characterization that involves conducting a series of tests on the material to determine its properties across multiple collaboration groups, and many more ``in-kind'' contributors who produce materials, services and models that are used by the materials science community. The term socio-scientific refers to the observation that in the big picture of open science ``it takes a village'' to produce and disseminate science. A material that is the end product of one lab may be a starting ingredient in the synthesis of an alloy by another lab. While one lab performs the synthesis of a material, another lab provides an instrument that measures and continuously monitors an experimental variable. Yet a third lab provides its infrastructure to perform specific characterization tests on the synthesized material. Away from the synthesis process, perhaps a Machine Learning lab provides a large language model (LLM) based service to identify related experiments from literature that the scientists would consult for materials discovery, experiment design and comparative analysis of material properties. Last but not the least, an under-resourced lab registered to the infrastructure will have access to any experimental process or resource (e.g., an instrument) captured within QDH to repeat and extend their science. We will present illustrations of this ecosystem in Sections \ref{sec:polystore} and \ref{sec:collab}.

\subsection{FAIR and UNIT: Design Pillars of QDH}
\label{sec:fairunit}
The QDH is designed to support the FAIR principles within the context of the materials synthesis ecosystem described above.
\begin{itemize}
    \item \textbf{Findability:} A human or machine user should be able to find, via a query interface or a web operation any resource or data related to any entity of the materials synthesis process, unless the owner of the information decides that the specific user is prohibited from such access. We discuss the implementation of findability in the next section.
    \item \textbf{Accessibility:} QDH enables the specification of a complex access control policy that combines role-based and discretionary access for any information and resource item (identified by a unique ID). More complex access control mechanisms like relationship-based access control \cite{bruns2012relationship,gupta2023path} can also be addressed when multi-organization ownership occurs. Access restrictions for a resource can be migrated and made completely public if desired.
    \item \textbf{Interoperability:} The QDH is designed for multiple organizations to share and contribute experimental information. In a later section, we elaborate the interoperability process in the context of new organization on-boarding. We will also demonstrate interoperability in the context of analytical service integration via the National Data Platform.
    \item \textbf{Reusability:} The QDH allows for synthesis procedures and experimental data to be reused by diverse users with access permissions. Templates for experimental and synthesis processes are made available along with related reuse governance and provenance tracking capabilities.
\end{itemize}
We supplement these FAIR principles by a new set of design criteria that we call UNIT. Like FAIR, UNIT criteria applies to the information component of the QDH and represents a user-centric view of the infrastructure.
\begin{itemize}
    \item \textbf{Usability:} QDH supports a wide variety of user groups ranging from new students to professional researchers to algorithms that interact with the infrastructure via different kinds of interfaces. These interfaces, developed through direct engagement with the user community are constantly refined and augmented. Further, the QDH maintains a log of system access that are analyzed to understand the utilization and utility of the system.
    \item \textbf{Navigability:} Navigability refers to the ability of a user to move from any information to any related information. While findability guarantees that every information or resource item can be discovered, navigability guarantees that if two information items are semantically related, the system will always provide a retrieval path from one item to the other.
    \item \textbf{Interpretability:} Interpretability is a design principle related to human users of the infrastructure and is deeply connected to the educational mission of the QDH. The principle states that any information access in the QDH can be supplemented by an additional variable that captures the user's professional level of technical literacy (e.g., high-school student vs. Ph.D. scholar), that enables the system to produce an explanation text in addition to the formal response.
    \item \textbf{Timeliness:} Unless specified otherwise, the system always prefers to respond to a request with the most recent data (e.g., the latest experiments) with the idea that they would be more valuable to the user to stay current with the content of the infrastructure. When avaiable, the prior versions of related data can be retrieved.
\end{itemize}

\subsection{GEMD++: Semantic Infrastructure of QDH }
\label{sec:gemd-ppod}
The socio-technical semantics of the QDH infrastructure significantly extends an existing model called GEMD \cite{GEMD} designed specifically for materials synthesis. GEMD \cite{GEMD}, originally, a JSON-based model, is designed to represent the detailed procedural steps of any materials synthesis process. In QDH, we adapt and extend the original GEMD model into a materials science collaboratory model - a graph data model, which is a directed graph where nodes and edges have types, and both nodes and edge can have a flat schema of (attribute, value) pairs. We note that similar models have been developed for other domains (e.g., PPOD \cite{gordon2021people} for environmental design).

\begin{table*}[ht]
    \caption{Semantic Node types for the GEMD++ model. \\In GEMD, a ``Run'' node has an outgoing edge to its corresponding ``Spec'' Node}
    \centering
    \begin{tabular}{|c|p{2.5in}|p{2in}|}\hline \hline
\multicolumn{1}{|c|}{\textbf{Node Type}} &\multicolumn{1}{|c|}{\textbf{Description}} & \multicolumn{1}{c|}{\textbf{Example}}\\ \hline \hline
\multicolumn{3}{|c|}{\textsc{GEMD COMPONENT}} \\ \hline \hline
Sample Root & Root for the GEMD representation of the synthesis graph of each sample & \\ \hline
Material Spec & Specification of a starting material for a synthesis experiment & Nickel in pellet form of purity between 90 and 95\%. A material may optionally have an ontological reference. \\ \hline
Material Run & Specification of a starting material for a specific instance of a synthesis experiment & Purchased Niobium with 94.3\% purity. It may have user-defined tags containing annotations \\ \hline
Ingredient Spec & An ingredient is a version of a material node that contains the \textit{amount or proportion} of the material used in a synthesis & Volume fraction range of the material going into an alloy\% \\ \hline
Ingredient Run & The actual proportion specified by absolute quantity, mass fraction, or volume fraction & 30\% by mass fraction from Alloy (Ni, T04) ingot \\ \hline
Process Spec & Name and description of a process along with process parameters & Grinding Synthesized EuS. A process may optionally have an ontological reference \\ \hline
Process Run & Exact process parameters and execution details &Arc Melting of alloy with parameters arc current/voltage, atmosphere and duration \\ \hline
Measurement Spec & Specific measurement performed during the experiment or later for characterization. If an instrument is used for the measurement, it will have an edge to the corresponding instrument node & A characterization for molecular composition of a synthesized material performed via X-Ray diffraction.\\ \hline
Measurement Run & Specific details of the measurement & Weigh the proportion of Vanadium in the alloy. This case does not have an associated file. But a characterization measurement run will have an associated measurement file produced by the instrument.\\ \hline
\multicolumn{3}{|c|}{\textsc{GEMD EXTENSION}} \\ \hline \hline
Instrument Run &An instrument only has a ``Run'' mode that specifies the instrument settings used for a specific measurement. An instrument node is connected to a measurement node via a ``uses'' edge & A Vibrating Sample Magnetometer specifies the magnetic field strength range used to determine the sample's magnetic hysteresis, permeability, and other properties \\ \hline \hline
Person &An individual person with edge to organizational entities and role. There is an optional edge from a GEMD process to a Person node indicating the role of the person in that project. & A UCSB Ph.D. student serving as a summmer intern at JHU runs some measurement operations.\\ \hline
Organization &Specifies the Institution and its subdivisions participating in a project &A collaboration project connects to multiple organizations. The edges from the organization node to the project node specifies the role of the organization in the project \\ \hline
Project &A temporary organizational structure for accomplishing a particular objective. Often a project is funded by one or more organizations. & The Quantum Foundry Project\\ \hline
Dataset &Any collection of data duly referenced to the creator, submitter. A dataset must specify the type of the data contained &A sample record, along with lab notebooks, image files, experimental procedure (GEMD) and Jupyter Notebooks is a QDH dataset \\ \hline
Report &A research report, a publication usually connected to a project or a set of synthesis experiments and to the authors of the report & \\ \hline
Tool & A codified technique for accomplishing a particular objective & A software library, a Jupyter Notebook that is used in performing any computation to produce a computed measurement \\ \hline
Service &A specific physical or computational facility that is offered in a standard manner for users &A custom Large Language Model (LLM) service that a scientist can use for material discovery \\ \hline
Infrastructure &A basic physical, organizational, or service facility (e.g., for computing or fabreication) needed for the operation of a project or a task & The Nautilus computation infrastructure used supporting services used by the QDH \\ \hline \hline
\end{tabular}
    \label{tab:node-types}
\end{table*}

\subsubsection{GEMD Component} In our adaptation, the graph-structured GEMD model is developed around the entities \textit{Materials}, \textit{Processes}, \textit{Instruments} and \textit{Measurements}, and their relationships (See Table \ref{tab:node-types}). Instruments can be used to conduct an experimental process (e.g., Chemical Vapor Deposition) or for characterization of material properties (e.g., Vibrating Sample Magnetometer). The task of materials synthesis starts with one or more materials, and applies a series of transformative processes under specific experimental settings to produce a final output material. For every synthesis instance, the GEMD model captures the end-to-end \textit{material history} of the output material. Table \ref{tab:node-types} shows some of the major node types supported in our framework. GEMD admits an elaborate scheme of attributes and values for its node types. For example, the temperature of a furnace can be modeled by the attribute schema:
\begin{verbatim}
    {   "type" : "uniform_real",
        "units" : "celsius",
        "lower_bound": 450.5,
        "upper_bound": 451.5  }
\end{verbatim}
which states the temperature in Celsius will have a value between the bounds and will be uniform through an experiment. 

Since GEMD needs to capture both experimental design and the execution of experiments, it uses separate constructs called \textit{Specification} (spec) and \textit{Run} for Materials, Processes and Measurements. A process specification, for example, specifies the \textit{permitted} environment (pressure, inert gases allowed) associated with the process, whereas a \textit{run} specifies a single actual instance of the process in one experiment. Very often, a spec can be associated with hundreds of runs where the parameters are varied on purpose to determine the end material's sensitivity to process variations.

\begin{figure}[t] 
\begin{mdframed}
\underline{\textsc{Relations}} (underlined attributes are primary keys)

\medskip

\textsc{samples}\texttt{(\underline{sample-id}, project-id, owner, date, [start-material-id], end-material-id, description, status)}\\
\textsc{materials}\texttt{(\underline{mat-id}, name, supplier, form, description)}\\
\textsc{material-prop}\texttt{(\underline{mat-id}, property-name, property-value)}\\
\textsc{measurements}\texttt{(\underline{measurement-id}, sample-id, material-id, instr-id, measure-date, measure-owner, measure-type, description, file-type, file-location-path)}\\
\textsc{instruments}\texttt{(\underline{instr-id}, type, make, model, specification)}\\

\underline{\textsc{Constraints}}

\medskip
\begin{enumerate}
\item \texttt{\textsc{samples}.sample-id = \textsc{measurements}.sample-id}
\item \texttt{\textsc{samples}.end-material-id = \textsc{measurements}.material-id}
\item \texttt{\textsc{samples}.end-material-id = \textsc{materials}.mat-id}
\item \texttt{\textsc{measurements}.material-id = \textsc{materials.mat-id}}
\item \texttt{memberOf(\textsc{samples}.start-material-id) = \textsc{materials}.mat-id}
\item \texttt{\textsc{measurements}.instr-id = \textsc{instruments}.instr-id}
\end{enumerate}
\end{mdframed}

\caption{A simplified version of relational schema maintained at QDH}
\label{fig:rel-schema}
\end{figure}
\setlength{\textfloatsep}{5pt}

We would like to make a few specific observations about our design around the GEMD model. 
\begin{enumerate}[label=(\alph*)]
    \item Each instance synthesized material (called a \textit{sample} in QDH) is associated with a singled GEMD graph, and $k$ repeats of the experiments (with or without parameter variation) results in a different sample with a distinct identifier, and $k$ corresponding GEMD graphs. Each GEMD graph corresponding to an experiment has a \textit{root node} that contains the \texttt{sampleID} of the corresponding sample.
    \item Not all information about materials, instruments and measurements is stored in the GEMD graph. For example, an instrument can be used in 100 different experiments -- so the details of the instrument are stored only once, in a relational table in a PostgreSQL DBMS. However, since the parameter settings of the instrument might be different for every run of every experiment, they are stored in the GEMD graph that is materialized in a Neo4J graph DBMS. If there is a publication with a set of experiments, then they are stored in a text collection indexed by the Apache Solr as well as via the QDrant vector database.
    \item Figure \ref{fig:rel-schema} shows a simplified the relational database schema maintained by the QDH for all of its material related data. Notice the constraints in the system maintained by the system to ensure referential integrity
\end{enumerate}

\subsubsection{GEMD Extension} The GEMD extensions shown in Table \ref{tab:node-types} represents the ``socio'' part of the techno-social framework for materials synthesis research in QDH, It identifies the entities connected with research efforts (e.g., people, organizations, infrastructure and services) to capture the research ecosystem. Adding this component to the original GEMD model lends itself to better handling of not only a more complete \textbf{research context} of multi-collaboration experimental research for quantum materials synthesis, but also it helps to identify \textbf{traceability and accessibility boundaries} related to this often-sensitive research domain. To illustrate the latter, suppose a project is trying use a material initially developed for a defense application as an ingredient for a new material that will have civilian use, the material history of the end product will trace back to its origin. When an authorized person in a Defense Lab looks at it, the entire process chain, including the Defense-internal subprocess will be visible but when a regular researcher views it, only the authorized subgraph of the network will be accessible. 

\subsection{Polystore-Based Storage and Implementation}
\label{sec:polystore}
The above-described variety of data stored in the QDH requires that data is distributed across relational databases, graph databases, text indices and file systems (specifically, SDSC's Cloud storage). To implement the design principles of findability and navigability, QDH uses a \textbf{polystore architecture}, which is a data mapping and query processing layer that sits on top of multiple storage, indexing and services to provide an integrated, semantic view of data. 

The user of the QDH system can enter through the QDH platform website and interactively create their sample records along with any associated information covered by the semantic types in Table \ref{tab:node-types}. Since the data is distributed, QDH maintains referential integrity across sites. For example, a sample record in the relational system has a verifiable file pointer for every binary data object deposited in the SWIFT cloud storage via a submission API. If the binary files are produced by some instrument, then the measurement node corresponding to the sample using this instrument also points to the same file to guarantee both findability and navigability criteria (Section \ref{sec:design}). In another case, a user may start by performing a ``bulk upload'' of data. For procedural data, this upload can be in the form a GEMD JSON file, or a graphML file that encodes the GEMD model. In the case the procedural data is very large, it can be uploaded as a directory of smaller JSON files, where each file represents one node of the large GEMD object, and the edges are implemented as file references inside the JSON object. In contrast, the system also allows a user to specify the experimental procedure via a graphical user interface called the \textbf{procedure editor}, which in turn produces a graphML object to be validated and ingested into the graph database. In the case where a user uploads a GEMD graph only, the system makes a best effort attempt to produce a normalized cross-store data set by ``shredding'' the GEMD graph to construct the relational records needed to preserve the polystore information schema.

\begin{figure}
    \centering
    \includegraphics[width=0.7\linewidth]{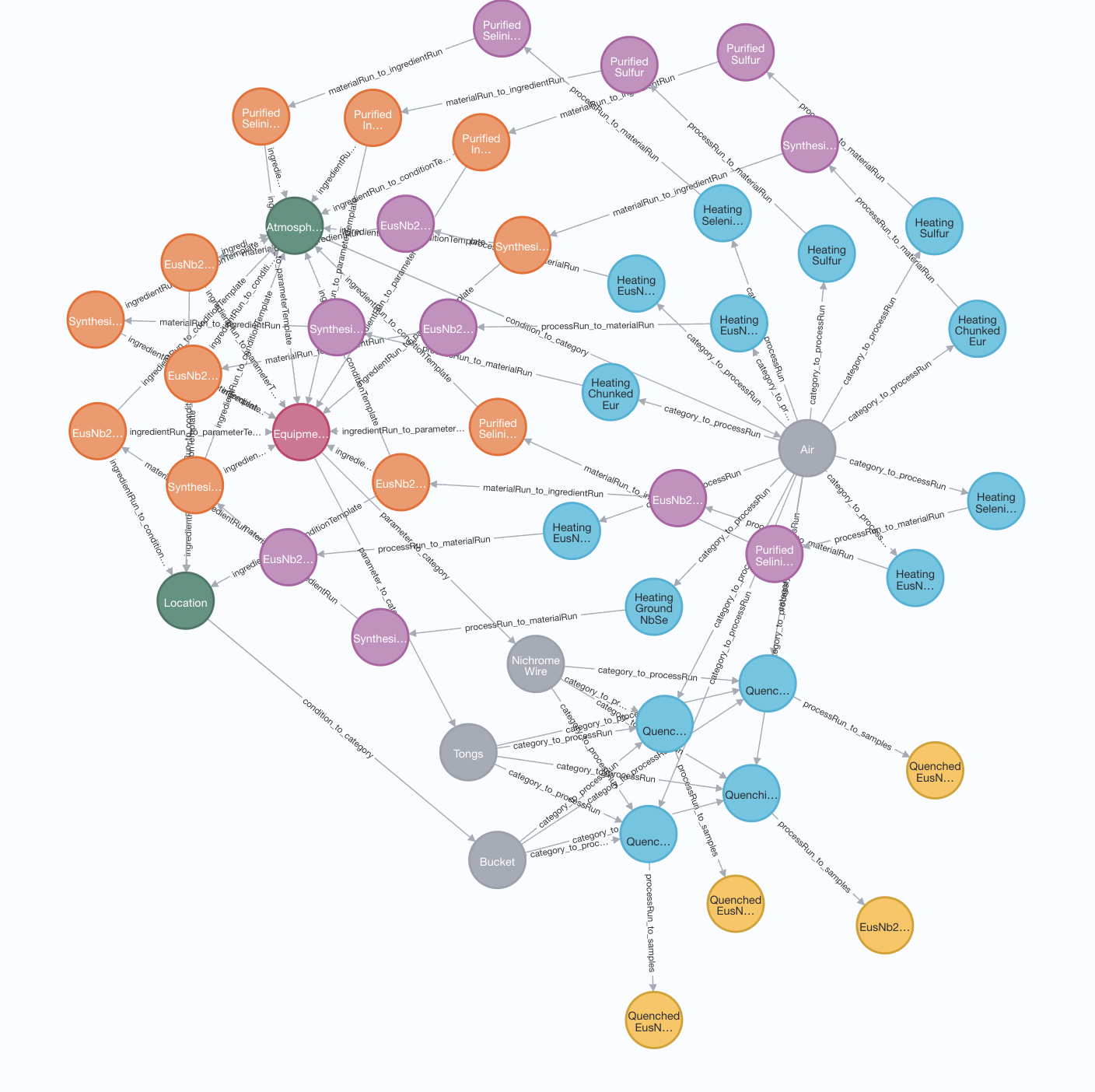}
    \caption{The material history of the sample specified by the query. The result has 4 samples (yellow), 15 process\_runs (blue), 11 material and ingredient nodes (light purple and orange) nodes. }
    \label{fig:mathistory}
\end{figure}

\noindent \textbf{Example 1: Querying the GEMD Material History Graph.} A student who wants to design a new material synthesis project queries the graph to identify synthesis processes that employs a heating process followed by a querying process.This can be accomplished via direct Cypher query to the Neo4J version of the graph.
\begin{mdframed}
    neo4j\_query = """ \\
			  match (n:process\_run)-[*]$\rightarrow$(m:process\_run)-[*]$\rightarrow$(k:samples)\\
			  where n =~ '.*Heating.*' and m.name =~ '.*Quenching.*'\\
			  return k.node\_id  """\\
neo4j\_result = graph.run(neo4j\_query)\\
\end{mdframed}
The result is the full material synthesis history of this sample as shown in Figure \ref{fig:mathistory}. The student can traverse the graph (i.e., from yellow to blue to purple and orange nodes) to understand the material synthesis process that was followed to create the node in yellow.

\noindent \textbf{Example 2: Cross-Database Queries.} A scientist wants to know which processes related to the starting material of Europa Sulfide requires heating. In the current version, the cross-model polystore query is written as follows.\\

\begin{mdframed}
\noindent x = cursor.execute("\textbf{select} *  \textbf{from} sample\_test  \textbf{where} sample\_name = 'Synthesized EuS';")\\

\noindent neo4j\_query = f"""\\
               \textbf{match} path=(n)-[*]$\rightarrow$(m)\\
               \textbf{where} n.sample\_id = "{x.sample\_id}" \textbf{and}\\
                     m.node\_id in {x.ending\_ids}\\
               \textbf{unwind} nodes(path) \textbf{as} node\\
               \textbf{with} node
               \textbf{where} node.type = 'process\_run' \textbf{and} node.name =~ '.*Heating.*'\\
               \textbf{return} node.node\_id \textbf{as} node\_id, node.name \textbf{as} name 
       """\\
    
\noindent neo4j\_result = graph.run(neo4j\_query)
\end{mdframed}
The above query finds samples with name ``Synthesized EuS", fetches its sample\_id and final materials (ending\_id) from the SQL subsystem, and then passes the list of (sample\_id, ending\_id) pairs to neo4j query to retrieve all the heating processes involved. The resulting list is:\\
\texttt{
'Heating Chunked Europium,Ground Purified Sulfur',
 'Heating EusNb2Se4 pellets (sealed vessel)',
 'Heating EusNb2Se4 vessel',
 'Heating Ground NbSe2 Mixture',
 'Heating Selenium',
 'Heating Sulfur'
}

\noindent \textbf{Example 3: Queries over Databases and File Systems.} A student seeks all X-Ray diffraction results from the cloud storage for samples that satisfy the conditions given in Example 1. The user issues the query:
\begin{mdframed}
query = SELECT *
FROM swift\_objects\\
WHERE swift\_objects.obj\_store\_path ~ (\\
    SELECT regex \\
    FROM characterization\_dictionary \\
    WHERE characterization = 'XRD' \\
) and sample\_id in '{neo4j\_result.sample\_id}';
\end{mdframed}
The regex table is a data dictionary that maps a data category like `XRD' to a corresponding file path pattern. The result of the query is a set of file paths stored in the cloud server (swift\_objects).

\begin{figure}
    \centering
    \includegraphics[width=\linewidth]{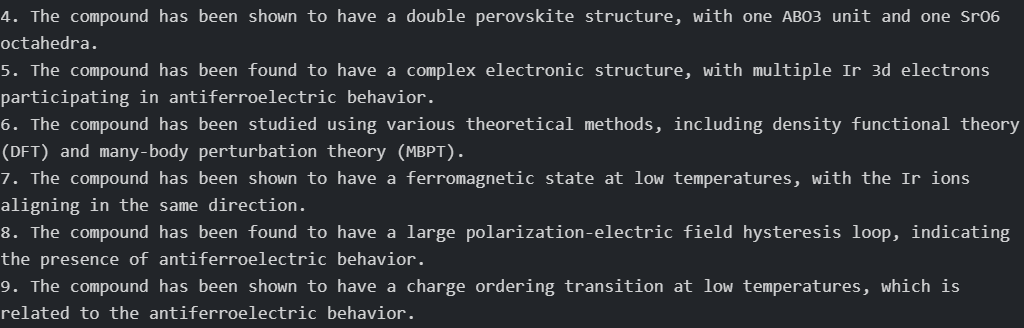}
     \caption{A screenshot showing a collection of discovered properties of $Sr_2IrO_4$ from multiple scientific papers.}
    \label{fig:LLMResponse}
\end{figure}

\noindent \textbf{Example 4: Querying Data and Discovery Services.} A materials researcher would like to discover the known properties of a material being considered for synthesizing an antiferroelectric alloy. To do this, the researcher uses the LLM service provided by the National Data Platform (NDP) as follows:
\begin{enumerate}
    \item The researcher, if authorized, uploads a collection of scientific publications to a private (or publicly visible) store. 
    \item She invokes an \textit{embedding service} that creates vector embeddings and places the resulting vectors in a vector store at NDP.
    \item The researcher queries the polystore to identify materials of interest (as shown previously).
    \item The list of materials are used in a prompt of the NDP LLM service to query the vector store.
\end{enumerate}
Figure \ref{fig:LLMResponse} shows a part of the response for a material called $Sr_2IrO_4$.

\subsection{Scalable Analysis Interface using Jupyterhub} 
\begin{figure}[!h]
    \centering
    \includegraphics[width=\linewidth]{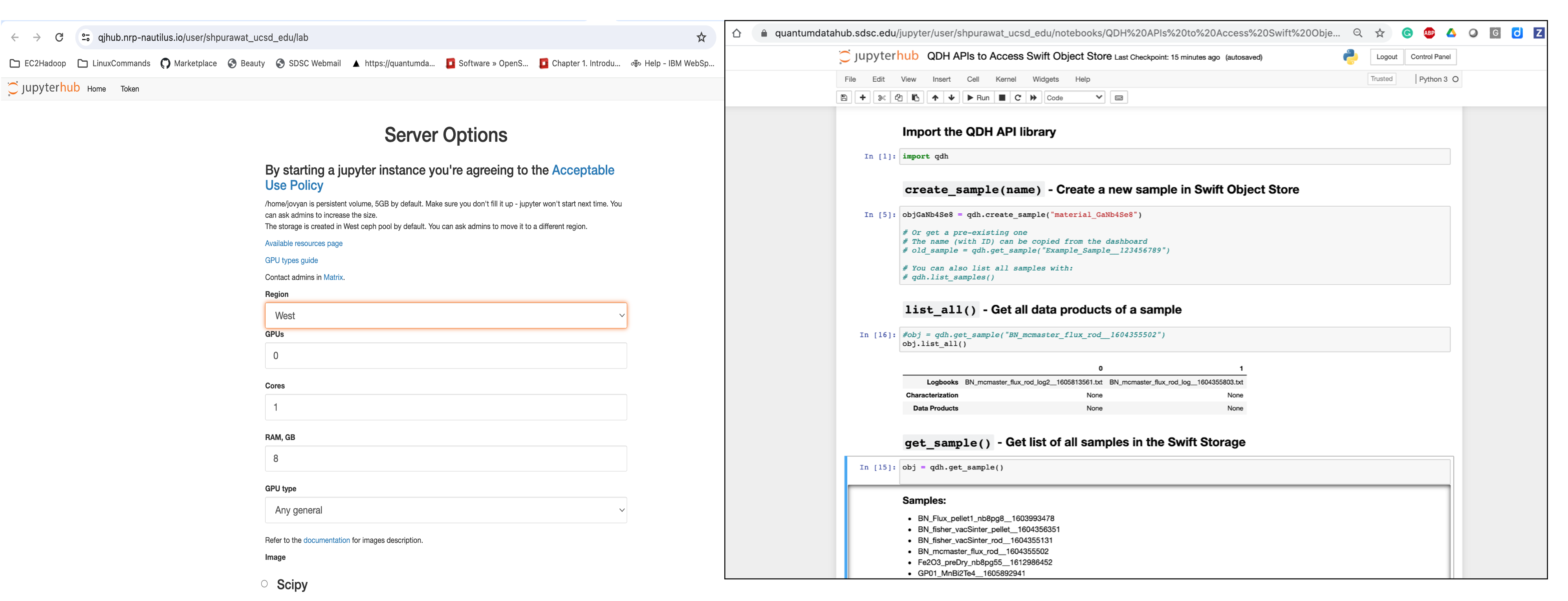}
    \caption{The left shows Jupyter Notebook server configuration interface and right side shows API calls to the get the data from QDH cloud storage to JN}
    \label{fig:jn}
\end{figure}

Accessibility and simplicity are core design principles behind the QDH Analysis Interface. The QDH interface can simultaneously support a large number of users and provide concurrent analysis environments through JupyterHub deployed on National Research Platform\footnote{https://nationalresearchplatform.org/} (NRP Nautilus). Users can dynamically choose the optimal combination of GPUs, CPUs, memory, and prebuilt software images for their Jupyter Notebooks, ensuring that computational resources are efficiently allocated. QDH APIs provide seamless access to Quantum Data objects on QDH Swift Cloud storage and empower users to build algorithmic solutions on centralized Jupyter notebooks. With its scalable infrastructure, flexible hardware selection, and API-driven data access, the QDH Analysis Interface provides user-friendly and efficient quantum data analysis. Figure~\ref{fig:jn} is a screenshot of the interface and a notebook showing API calls for data movement to and from QDH cloud storage. 

\section{Facilitating Science and Collaborations}
\label{sec:collab}

\subsection{Facilitating Science} 
\label{sec:facilitating}

The Quantum Data Hub facilitates advanced material science research by providing powerful features for data management, searching, analysis, collaboration, and sharing. This section presents a few cases that show how QDH features enhance material science research and development.
\begin{itemize}

\item \noindent \textit{ \textbf{QDH Procedure Editor facilitates Digital Documentation of Material Synthesis Processes:}}
A web-based interface enables researchers to document their lab notebook processes efficiently in a workflow graph format, as shown in Figure~\ref{fig:qdhpe}. The Procedure Editor features a graphical user interface with drag-and-drop functionality, simplifying the design and documentation of material synthesis processes. The QDH Procedure Editor includes a library of standard process and measurement nodes that support various material sample types, such as poly-crystal, single-crystal, thin-film, and devices.
This enhances material science by allowing the documentation of both successful and failed material synthesis attempts. In the future, we aim to use machine learning to extract knowledge and patterns from this data, providing researchers with valuable guidance for synthesizing materials.

\begin{figure}[!h]
    \centering
    \includegraphics[width=\linewidth]{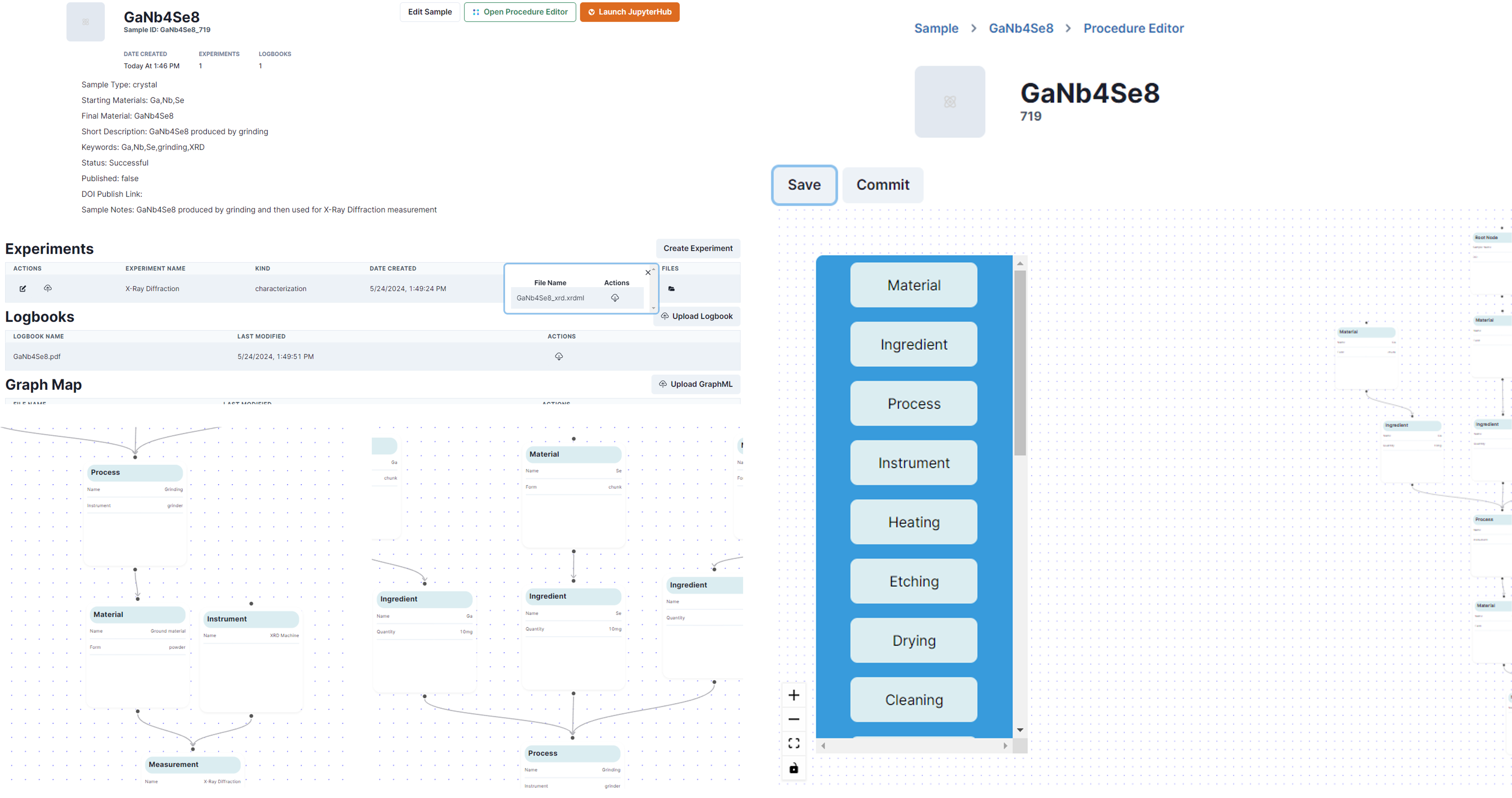}
    \caption{The figure shows GaNb4Se8 Sample metadata and synthesis graph stored created in QDH Procedure Editor Interface}
    \label{fig:qdhpe}
\end{figure}

\item \noindent \textit{ \textbf{QDH facilitates Reproducibility and Collaboration through Standardized Data Sharing and Ingestion:}}
The standardized GEMD Data Model allows for consistent logging of material synthesis processes. This ensures easy data sharing and ingestion from different groups and instruments, facilitating reproducibility and collaboration.

\item \noindent \textit{ \textbf{QDH facilitates a Collaborative Data Analysis Environment:}} 
Multiple users can access the Jupyter Notebook environment through JupyterHub on NRP Nautilus. This setup streamlines data analysis and provides easy access to QDH Cloud Storage.

\item \noindent \textit{ \textbf{QDH facilitates automated Large Scale Ingestion of Samples and Material Synthesis Processes through Bulk Upload:}}
The system supports the bulk upload of thousands of samples, enabling efficient large-scale data ingestion. Samples can be ingested from literature using natural language processing (NLP) and from instruments, automating the data collection process and expanding the dataset significantly.
\item \noindent \textit{ \textbf{QDH Supports Collaboration among Multiple Foundries and Groups through Advance Access Control:}}
QDH offers group-based access, allowing members of the same group to access shared samples, which promotes internal collaboration. Additionally, researchers can use discretionary access control features to share and collaborate with users from different groups or foundries, enhancing cross-group and cross-institution collaboration. This feature is explained in detail in Section \ref{sec:access}.
\item \noindent \textit{ \textbf{QDH Facilitates Easy Accessibility to Material Synthesis Data:}}
QDH uses Polystore-Based Storage and an Advanced Query Processing layer to ensure easy access to material synthesis data. It supports the FAIR (Findable, Accessible, Interoperable, Reusable) data principles, as discussed in Section \ref {sec:design}, making data well-organized and easily accessible. This improves data management, sharing, and reuse, thereby accelerating scientific discovery.
\end{itemize}

Integrating these features, the QDH provides efficient tools for material science research that will lead to more rapid and innovative material discoveries.

\subsection{Adding New Collaborations}
\label{sec:extensibility}
The QDH is designed to be a flexible and extensible infrastructure so that it can enable new collaborations. We present two cases -- the first represents the technical onboarding of a new research group into the QDH ecosystem, and the second represents an indirect collaboration with a service infrastructure as alluded to in \ref{sec:design}. 

\noindent \textbf{Technical Onboarding:} When a new research group joins the QDH, their materials synthesis process is different from the currently supported schema of the QDH. We present an actual use case where a new group (MonArk) joined the QDH for storing data for 2D crystallites, 2D heterostructures and 2D devices. Further, they specify a new relational schema structure detailing the new tables. Their conceptual model has an aggregation hierarchy where devices are composed of heterostructures which are composed of crystallites that are derived from bulk crystals -- a structure that does not exist in the original QDH schema (see Figure~\ref{fig:rel-schema}). Further, they have identified new types of characterization data (e.g., gate-dependent responses and field-dependent transport data for devices) that will be stored as files in the cloud storage. To ensure that these additions do not contradict but extend the existing schema, we create a set of \textit{schema reconciliation mappings}, and a set of \textit{data dictionary updates}. 
\begin{itemize}
    \item \textbf{Schema Reconciliation:} We add the new tables to the QDH schema However, we add the identifiers of each table of this new schema to a \textbf{federated schema} where we construct and enhanced sample entity that is a logical union of the existing samples entity and the new tables. We also construct a semantic type for each table such that a query like ``find all samples of 2D devices'' can be directed to the appropriate tables.
    \item \textbf{Data Dictionary Update:} The system maintains a data dictionary for the different characterization methods and the property they characterize (e.g., XRD files characterize X-Ray crystallography data and measures material composition). It also has a regular expression that can be used to identify this type of data from the cloud services. The new characterization data updates this dictionary based on the new user's input.
\end{itemize}

\noindent \textbf{External Services:} A scientist using QDH can indirectly collaborate with an external group that provides a computational service. Recall our query Example 4 in Section \ref{sec:polystore}, where a suite of LLM services provided by the National Data Platform is used by a QDH scientist. To enable such a cross-facility collaboration, both groups must share a common administrative domain of access control, such that an API token generated by the domain can be cross-used across both infrastructures. In our case, access tokens issued by the two CILogon based systems is utilized. We view this as a generic mechanism of service infrastructure that can be used by open science projects.

\section{Facilitate sharing of data and materials synthesis procedure through Access Control}
\label{sec:access}


\begin{figure}[ht] 
\centering
    \includegraphics[width=\linewidth]{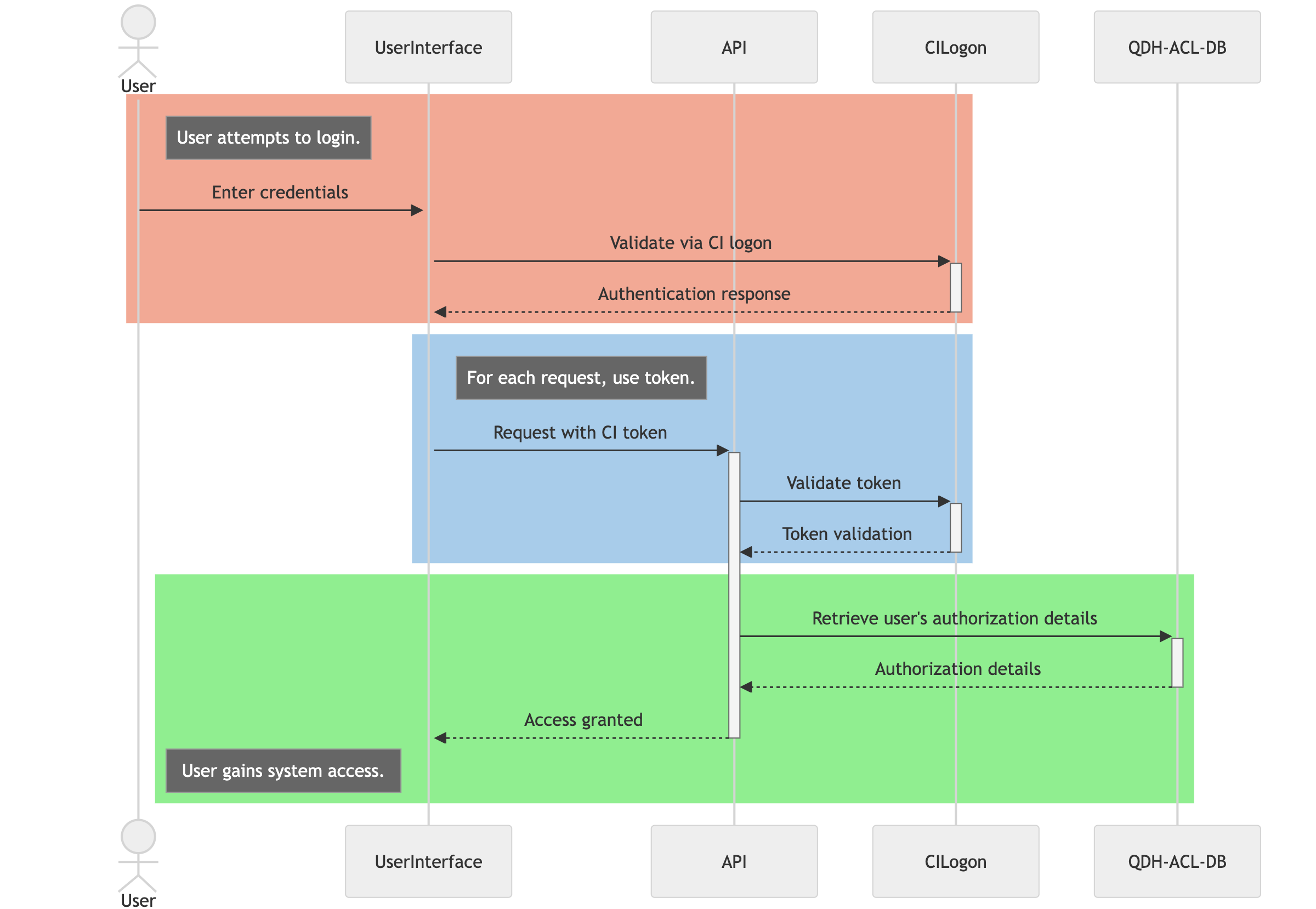}
        \caption{QDH Authentication Protocol sequence diagram illustrating the Security Protocols.}
    \label{fig:auth}
\end{figure}

The QHD system is designed to offer robust data security and flexible access control for users in material science and material synthesis. Additionally, it was developed to facilitate the sharing of material synthesis processes, data, notebooks, and lab reports, thereby fostering collaboration among various groups. Having the primary focus was on material science and synthesis, the system is versatile enough to be applied to any general experimental scientific framework.

In this current version,  for authentication and authorization, QDH employs a token-based CILogon authentication system as illustrated in Figure \ref{fig:auth}. Initially, the user interface(UI) validates authentication through the CI log-in. Following the initial log-in, it forwards every request to the back-end using the CI log-in token. The API then validates this token by making a secondary call to the CILogon provider. The authorization or access control database is integrated into our system. Upon successful user authentication, the system fetches the user's authorization information from this database. Figure \ref{fig:auth} illustrates the various layers of the system through a sequence diagram. 

\textbf{Authorization Model.} The authorization model is responsible for providing access to the object. In QDH, we implemented a group-based user authorization model with discretionary access control. The authorization components in this model are defined by the following functions\cite{matt2002computer}:
\begin{itemize}
    \item For any instance of authorization $i$, we can write :
\begin{equation}
    A_i \longrightarrow f(O_i, S_i, \beta_i)
    \label{eq:myEquation}
\end{equation}
    \item The total authorization for a user could be written $Z = A[1..n]$
\begin{itemize}
    \item \textbf{Object$(O_i)$:} The Object of the authorization model is samples and their related artifacts, e.g., measurements and logbooks. A collection of samples can be marked as public, granting access to any member of the QDH. Except for the objects available as public samples, all other objects belong to the research group that produced them. We consider the sample as the primary object, which may have several associated artifacts. If a subject has access to the object, they can read, write, and update any components linked to the object. Normally, deleting is not a permissible action in QDH. For example, if a logbook is linked to a sample and a subject has access to that sample, the subject will inherently have full access rights to the logbook.
    \item \textbf{Subject$(S_i)$:} In QDH, subjects are members of research groups, and each subject must belong to at least one group within the system. However, within these groups, the subject can assume one role from the available roles, and the group owner is responsible for assigning that role. These roles are discrete, and there is no information or security propagation implied. The system includes groups for PIs, researchers, Ph.D. students, and students. The PI acts as the owner of the group and has the authority to add members to their group. A group member can access any object accessible to their assigned group and perform all operations on it. In the current implementation, a subject can belong to only one group. 
    \item \textbf{Access Rights $(\beta_i)$:} Each member of a group has full access to their respective group, and the access rights include read, write, and update permission. As we mentioned, in order to maintain academic and research integrity, the delete operation is not available to the group members. However, the system is capable of performing restore and delete actions for exceptional scenarios. The QDH access control system exclusively considers positive access rights $(+ve)$, and operates under a closed system assumption\cite{samarati2000access}. If a subject has positive access rights to an object, those rights are valid. The system does not accommodate any form of negative access rights$(-ve)$. Furthermore, the system allows collaborations between groups by allowing an additional discretionary access control primitives. A group owner or a designated representative can invite a member from another group through a discretionary rights, and the subject must be explicitly assigned to all objects in the foreign group. The discretionary access component is individually and explicitly verified for each subject, ensuring that permissions do not extend to the subject's original group. Therefore, any access rights granted to a user through the dictionary access control do not automatically propagate to their group.
    \item \textbf{Access Conflict:} QDH addresses access conflicts by independently evaluating group-based policies and discretionary policies. Therefore, the access matrix $Z$ for an individual is composed of two distinct components: $Z_g$ for group-based components and $Z_d$ for discretionary components.  Hence, we could say that:
    $$Z \rightarrow Z_g + Z_d $$
Additionally, since negative access control ($-ve$) is not considered, and combining the equation \ref{eq:myEquation} we could say 
\begin{equation}
    \forall i, (\mathbf{\beta})_i \neq (-ve) \implies \forall i, (\mathbf{Z}_d)_i = (+ve)
\end{equation}

Thus, $Z_d$ will invariably be positive$(+ve)$ as $\beta$ could not be negative, which leads us to conclude that there will be no conflict in QDH \cite{dasgupta2012graph}. 
\end{itemize}
\end{itemize}
\section{Conclusion and Future Work}

This paper described the design principles, data system and AI integration components behind the Quantum Data Hub to enable collaborations across diverse users across labs. We will further extend the GEMD++ models to incorporate other aspects of Quantum Materials Science. We can enhance the system by improving the authorization mechanism to allow for more detailed access management. At present, the sample is considered to be the atomic unit of the object, but it can be further divided. Access control mechanisms could be applied to specific parts of the sample. For example, if a particular methodology is sensitive, a user could obscure that specific method while allowing access to others. The negative access specification is a forward-looking concept. At present, within a group, it is not possible to assign negative access, but this could be a significant feature in the future. Currently, the system lacks an audit trail feature, which could be introduced as an additional capability. This feature could provide users with constructive feedback by informing them about their actions. The QDH user interface could be further modified and updated in the future to make it more centered around different user personas, making it more user friendly. We will also extend the query and analysis facility of the system to use local and external services for public and protected data.

\section*{Acknowledgment}

The authors would like to thank the Quantum Foundry (https://quantumfoundry.ucsb.edu/) and WorDS teams for their collaboration and support of this study by NSF 1906325, 1906383, 1909875,and 2333609 for National Data Platform, and  the Nautilus Kubernetes Cluster of the National Research Platform 2112167 and 2120019.

\bibliographystyle{IEEETran}
\bibliography{QDH}

\end{document}